\documentstyle[prd,aps]{revtex}
\begin{document}
\def\la{\mathrel{\mathchoice {\vcenter{\offinterlineskip\halign{\hfil
$\displaystyle##$\hfil\cr<\cr\sim\cr}}}
{\vcenter{\offinterlineskip\halign{\hfil$\textstyle##$\hfil\cr<\cr\sim\cr}}}
{\vcenter{\offinterlineskip\halign{\hfil$\scriptstyle##$\hfil\cr<\cr\sim\cr}}}
{\vcenter{\offinterlineskip\halign{\hfil$\scriptscriptstyle##$\hfil\cr<\cr\sim\cr}}}}}

\def\ga{\mathrel{\mathchoice {\vcenter{\offinterlineskip\halign{\hfil
$\displaystyle##$\hfil\cr>\cr\sim\cr}}}
{\vcenter{\offinterlineskip\halign{\hfil$\textstyle##$\hfil\cr>\cr\sim\cr}}}
{\vcenter{\offinterlineskip\halign{\hfil$\scriptstyle##$\hfil\cr>\cr\sim\cr}}}
{\vcenter{\offinterlineskip\halign{\hfil$\scriptscriptstyle##$\hfil\cr>\cr\sim\cr}}}}}

\title{Conceptual issues in combining general relativity and
quantum theory}
\author{T. Padmanabhan$^{1}$\thanks{E-Mail:~paddy@iucaa.ernet.in}\\ 
 Inter-University Centre for Astronomy \& Astrophysics\\
Post Bag 4, Ganeshkhind,\\
Pune - 411 007, India \\
}
 
\maketitle
 
\begin{abstract}

\noindent Points of conflict between the principles of general relativity and
quantum theory are highlighted. I argue that the current language of
QFT is inadequete to deal with gravity and review attempts
to identify some of the features which are
likely to present in the correct theory of quantum gravity.

\end{abstract}

\section {Introduction}
The question of bringing together the principles of quantum theory
and gravity deserves to be called ``the problem'' of theoretical physics
today. The history of failures in this attempt illustrates not only the conceptual complexity of the problem but also the sociology of science
in the late twentieth century. Since 
Jayant will be sympathetic to  my --- rather heretical ---  way of thinking about this issue,
I thought a description of my views on this subject will be appropriate
for this volume.

\section{ The miracle of quantum field theory}

In proceeding from classical mechanics [with finite number of
degrees of freedom] to quantum mechanics, one attributes
operator status to various dynamical variables and imposes the commutation relations among them. These relations, and an expression for the hamiltonian
operator $\hat H(\hat q,\hat p)$, allow  us to write down the equations for the
time evolution of the operators. If these equations can be solved, then we
can determine the full structure of the theory. Often, it is
conveneient to provide a repesentation for the operators in terms of normal
differential operators so that the problem can be mapped to
solving a partial differential equation --- say, the time-dependent Schrodinger equation --- with specific boundary conditions. Such problems are mathematically well defined and tractable, allowing us to  construct
a well defined [though, in general, not unique] quantum theory for a classical
system with finite number of degrees of freedom.

The geralisation of such a procedure to a {\it field} with infinite number of
degrees of freedom is {\it not} straightforward and is fraught with 
conceptual and mathematical problems. Given a classical field with some dynamical variables,  one can attempt to quantise
the system  by elevating the status of dynamical variables to operators
and imposing the commutation rules.
But finding a well defined and meaningful representation for this commutator algebra is a nontrivial task. Further, if one tries to extend the approach
of quantum mechanics [based on Schrodinger picture] to the field, one obtains
a {\it functional } differential equation instead of a partial differential
equation. The properties  --- let alone solutions! --- of this equation are
not well understood for any field with nontrivial interactions. Somewhat
simpler (and better) approach will be to use the Heisenberg picture and try to
solve for the operator valued distributions representing the various dynamical
variables. Even in this case, one does not have a systematic mathematical
machinary to solve these equations for an interacting field theory. The
procedure to quantise an arbitrary [but well defined] classical field theory
fails right at the outset due to inadequete mathematical apparatus. We
have no right to expect  quantum field theories to exist!

It is, therefore, quite surprising to me that quantum field theories indeed
could be developed and used to make veriafiable predictions. To see how
this miracle was achieved, let us look at the prototype of quantum field theory,
viz. QED. The evolution equations for operators in QED [in 3+1 dimensions] cannot be solved exactly; however, it is possible to set up a perturbation expansion for
these variables in powers of the coupling constant $(e^2/\hbar c)\approx 10^{-2}$. The lowest order of the perturbation series, in which all interactions
are switched off,  defines the so called {\it free field} theory. It is possible
to map this theory to one describing infinite number of noninteracting
harmonic oscillators and solve for the dynamics of any one of the oscillators completely. The
perturbation expansion can be then used to obtain the ``corrections''
to this free
field theory. Several issues crop up when such an attempt is made:

(a) To begin with, the decomposition of the field in terms of the harmonic
oscillators is not unique and there exists infinite number of inequivalent
representations of the basic commutator algebra for the system. 
This shows that 
``physical'' quantities like ground state, particle number etc. will depend on
the specific representation chosen and will not be unique.

(b) Since the system has infinite number of degrees of freedom, quantities
like total energy can diverge. The actual form of the
divergence depends on the representation chosen for the algebra and the
differences between infinite quanties may retain a representation dependent
[finite] value, unless one is careful in regularising such expressions. In
some cases, one may be forced to choose particular set of harmonic oscillators
because of the boundary conditions. Then, the difference between
two infinite quantities could  be physically relevant (and even observable as in the case of, for example, Caisimir effect).

(c) The situation becomes worse when the perturbation is switched on. In general,
the perturbation series will not converge and has to be interpreted as an
asymptotic expansion. Further, the individual terms in the perturbation series
will not, in general, be finite creating a far more serious problem. This arises
because the amplitude for propagation of a free field quanta,  of mass $m$ and
 euclidian  momentum $p$ varies as $(p^2 +m^2)^{-1}$, which does not die down sufficiently fast at large $p$. This, in turn, is related to the fact that virtual quanta
of {\it arbitrarily} high energy are allowed to exist in the theory [incorporating Lorentz invariance at arbitarirly small length scales] and
still propagate as free fields.

(d) Perturbation theory completely misses all effects which are nonanalytic
in the coupling constant. In QED, for example, perturbation theory cannot
lead the result that an external electromagnetic field can produce
$e^+-e^-$ pairs, since this effect has nonanalytic dependecy on $e$. \cite {schwing1951}   One cannot even estimate the seriousness of this problem since very few nonperturbative
results are known.

How does one cope up with these difficulties? Issue (a) is handled by choosing
one particular representation for the free field theory by fiat, and working
with it --- and ignoring all other representations which are not unitarily
equivalent to the same. This also dodges the issue (b) provided some means of
regularisation can be found to discard the infinities of the free field theory.
Once a representation for the harmonic oscillators is chosen, this can be
implemented by a procedure like normal ordering. Issue (d) is accepted
as a failure of the method [at least by the honest researchers!] and then
ignored. Most of the successful effort was concentrated on handling the
problem of infinities {\it in the individual terms} of the perturbation series,
that is, on issue (c). The pardigm for handling these infinities can be
stated in terms of 
the concept of renomalization which, by itself, has nothing to do
with any    divergence. In the simplest terms, renormalization 
expresses the fact that  the interactions will   change
the values of the various coupling constants in the theory; that is, the 
physically observed coupling constants are the ``renormalized'' ones
and not the ``bare'' ones which appear in the original Lagrangian.
 The phenomena of renormalization
exists, for example, in condensed matter theories where both the bare
and renormalized coupling constants can be finite. In the context of field 
theory, renormalization can provide a means to eliminate divergences,
 if {\it all} the divergent terms of a perturbation expansion can be
eliminated by redefining the coupling constants in the theory. For an
arbitrary  field theory, we have no assurance that all the divergences can
be so eliminated; in fact, it is quite easy to construct well defined classical field theories for which divergences cannot be eliminated by this process. 

The unexplained
miracle of 20th century quantum field theory lies in the fact that several
physically relevant field theories --- describing quantum electrodynamics,
electro-weak interactions and QCD --- belong to this special class of {\it perturbatively renormalisable} theories. For such theories, perturbation
series can be developed as an algorithmic procedure  to evaluate matrix elements for transitions between  asymptotic states of the free field theory, to any order 
in perturbation theory. The agreement of such predictions with 
observations led to (several nobel prizes  and)  a religious 
faith in perturbative renomalization as the paradigm of quantum
field theory by late 60's - early 70's.
Nobody knows why this mathematically non-rigorous,
conceptually ill-defined, formalism of perturbative quantum field theory
works. The miracle becomes even more curious when we notice that
the bag of tricks fail miserably in the case of gravity.

\section{The edifice of general relativity}

Until early seventies, most of the hardcore particle physicists used 
to ignore general relativity and gravitation and the first concrete attempts in
putting together  principles of quantum theory and gravity were led by 
general relativists (see e.g. ref.\cite {dwitt67}). It was clear, right from the beginning, that
this is going to be a formidable task since the two ``theories of principle''
differed drastically in many aspects. The key features of gravity
which are of relevence in this context are the following:

(a) The Lagrangian describing classical gravity, treated as a function of
$h_{ik}=g_{ik}-\eta_{ik}$, is {\it not} perturbatively renormalizable;
in fact, there does not exist any simple redefinition of the field variables
which will lead to a perturbatively renormalizable theory. So the most
straight forward approach, based on the belief that nature will continue to be kind
to us, is blocked. The miracle fails.

(b) The principle of equivalence implies that any resonable description of
gravity will have a geometrical structure and that gravitational field
will affect the spacetime intervals in a specific manner. This inescapable
conclusion leads to several corollaries, all of which make gravity 
an odd-man-out: (i) To begin with, this makes the spacetime itself a dynamical 
entity and not something which can be prescribed beforehand. (ii) Secondly,
the description of gravitational field in terms of the metric 
tensor $g_{ik}$ translates into a {\it constrained} dynamical system; that is,
the true degrees of freedom of gravity are only 2 per event rather
than the full set of 10 functions contained in $g_{ik}$. Understanding 
the nature of constraints in general relativity --- and implementing it
in different descriptions of quantum theory --- turn out to be a very non trivial
task. (iii) Thirdly, the geometrical description leads
 to a fairly unique (class 
of) Lagrangian(s) for the gravitational field. The equivalent Hamiltonian formulation
of the theory in terms of 3-geometries lead to a degree of freedom (conformal factor) which is unbounded from below. 
(iv) The geometrical structure also implies that there is no prefered coordinate
system in the presence of gravitational field. In fact, there is no unique
and meaningful separation of the various effects as those due to gravity
and those due to noninertial forces, if we stick to the metric
tensor as the fundamental physical variable. For a general gravitational
field, there will be no way of choosing a special class of spacelike hypersurfaces
or a time coordinate. 

(c) Gravity affects the light signals and hence determines the causal 
structure of spacetime. In particular, gravity is capable of generating
regions of spacetime from which no information can reach the outside
world through classical propagation of signals. This feature, which 
may be loosely called `the existence of trapped surfaces' has no 
parallel in any other interaction. 

(d) Since all matter gravitates, the gravitational field becomes more and more
dominant at larger and larger scales. In the limit, the asymptotic structure of
spacetime is determined by global, smoothed out distribution of matter 
in the cosmological context. In such a case, the spacetime will not be
asymptotically flat in the spatial variables at any given time. The behaviour
of the spacetime for $t\to \pm \infty$ will also be highly
non-trivial and could be dominated by very strong gravitational fields. 

(e) All energies gravitate thereby removing the ambiguity in the zero level
for the energy, which exists in non-gravitational interactions. This feature also suggests that there is no such thing as a free, non-interacting
field. Any non trivial classical field configuration will possess certain amount
of energy which will curve the spacetime, thereby coupling the field to itself
indirectly. Gravitational field is not only nonlinear in its own coupling,
but also makes {\it all other fields} self-interacting.

(f) The coupling constant governing gravitational interaction has a non trivial dimension in the language of quantum field theory; $E_P\equiv (G/\hbar c)^{-1/2}$ has the 
dimensions of energy in contrast to $(e^2/\hbar c)$ which is dimensionless.
Simple power counting arguments based on this result will 
show that gravity will be perturbatively non renormalizable. Further,
one can construct a quantity with dimensions of length, $L_P \equiv (G\hbar /c^3)^{1/2} \approx 10^{-33}$ cm, from the gravitational coupling constant.
Though no formal proof exist, it is very likely that quantum gravitational effects will modify the spacetime structure at length scales comparable to $L\approx L_P$. In fact, simple thought experiments combining the principles
of quantum theory and gravity show that the planck length acts as a `zero-point-
length' to any spacetime.
 (see e.g. ref \cite{tp87}) Any correct formulation of quantum gravity must
have the infrastructure to incorporate this feature just as the operator
description of quantum mechanics is capable of incorporating the uncertainty
principle.

(g) The truely remarkable feature of classical general relativity is that
{\it this theory is fundamentally wrong}. This is most easily seen
from the fact that one can ask questions ---  in the form of thought experiments --- to which the theory cannot
provide sensible answers. One such question could be the following:
``A neutron star of mass $6 M_\odot$ collapses to form a blackhole. How will
the physical phenomena appear with respect to a hypothetical observer
on the surface of the neutron star at arbitrarily late times
as measured by the observer's clock?'' Such  questions cannot be answered in classical general relativity because the relevant equations lead to an infinte curvature singularity. Such a theory must clearly be wrong and has to be replaced by a 
better formulation at very strong curvatures. 

The   features (a) to (d)  already suggest that there are fundamental contradictions
between the formulation of quantum field theory and that of general relativity.
Given the result (a), one could have taken two separate routes: (i) How can
gravity be made to conform to the tenets of QFT ? or (ii) Why did QFT
work in the case of other interactions and how should QFT be modified
to handle gravity ? Historically, most of the effort went into route (i)
and led to a blazing trail of failures. This is in spite of the fact that
many of the features listed above show that
contradictions of language surface even when one tries to develop a quantum 
field theory in an external gravitational field (without worrying 
about the quantization of gravity itself).  Since gravity does not allow
a prefered slicing of the spacetime, quantum field theory needs to be 
formulated without using any prefered representation for the operator 
algebra. Loosely speaking, this implies that there is no generally covariant  definition for the vacuum state (or particle excitations)
in a generic curved spacetime. Infinite number of inequalent representations
exist and we have no means of choosing any one of them as `more physical' 
than another. It is clear that
such a description --- based on a ground state and the particle-like excitations ---
is of very limited value and will not survive the transition to the next layer,
say, the one in which we need to take the back reaction of the particle production into account. 

An abstract way of stating the same conclusion is as follows: Gravity is
inherently local (local coordinate charts, observers, freely falling frames ...)
while the standard formulation of QFT is global (global spacelike hypersurface,
global mode functions, ....). There is no such thing as `one-particle-state-
at-the-event-${\cal P}$' in QFT and there are serious problems in defining
any such concept. 

More  difficulties arise from the 
feature (c) listed above. When gravity makes certain regions inaccessible, the 
data regarding quantum fields in these regions can ``get lost''. This requires
reformulation of the equations of quantum field theory, possibly by tracing 
over the information which resides in the inaccessible regions --- something
which is not easy to do either mathematically or conceptually. Trapped surfaces
also highlight the role of boundary conditions in  QFT. The structure of a free
field propagating in an arbitrary spacetime can be completely specified
in terms of, say, the Feynman Greens function $G_F(x,y)$ which satisfies
a local, hyperbolic, inhomogeneous, partial differential equation. Each solution to this equation provides a particular realization of the theory. In other 
words, there exists a mapping between the realizations of the 
quantum field theory and the relevant boundary conditions to this equation which specify a useful solution. When trapped surfaces exists, the differential operator governing the Greens function will be singular on these surfaces 
(in some coordinate chart) and the issue of boundary conditions become far more complex. It is, nevertheless possible --- at least in simple cases with compact trapped surfaces --- to provide
an one-to-one correspondence between the ground state of the theory and the 
boundary conditions for $G_F$ on the compact trapped surface. In fact,
the Greens function connecting events outside the trapped surface can be completely determined in terms of a suitable boundary condition on the trapped surface, indicating that
trapped surfaces acquire a life of their own even in the context of QFT in CST.
In a way, the procedure is reminiscent of renormalisation group approach,
but now used in real space to integrate out information inside the trapped
surface and possibly replace it by some boundary condition. 

In this
connection, it is worth noting that effects like particle production by
a blackhole (or expanding universe) are {\it infrared} phenomena and arises
due to the coupling of modes at large scales. [The conflict between local
 GR and global QFT is again apparent]. The ultraviolet modes are
comparitively local and decoupled. This is somewhat different from standard
situations in QFT where the ultraviolet modes get coupled due to interaction
and the infrared ones get a free ride. Integrating out the information inside
a trapped surface in real space might also translate into a renormalisation group
approach in fourier space {\it with infrared modes integrated out}.

The importance of cosmological solutions in classical
gravity [item (d)] led to the investigations in quantum cosmology and the possibilities
of `wave function of the universe'. Two features emerged from these
attempts: (i) It may be possible to circumvent the classical cosmological 
singularity in quantum cosmological models. (ii) If the ground state of
the universe is globally determined, the boundary conditions could
also lead to specification of the ground state for matter fields, thereby
providing a quantum version of Mach's principle \cite {tp90}. Both these results are
tentative and nonrigorous but go  to show the richness of possibilities. The feature (d), however,  creates  problems in formulating quantum
field theory in terms of scattering amplitudes or asymptotic ``in'' ``out'' states. Such concepts are meaningful when the global spacetime structure is 
externally specified but not when dynamics determines the structure of asymptotic universe. 

 I think the  key {\it physical} message from some of these
investigations is  the following: Fields are more important than particles and could
be more robust entities. 
In fact, this conclusion is apparent even from the existence of 
a phenomena like Casmir effect which cannot be explained in terms
of virtual particles and is independent of the coupling constant $(e^2/\hbar c)$ of the 
perturbative theory.
{\it This is in sharp contradiction with 
the phisolophy of perturbative gauge theories in which the particle physicist
uses fields   just as a tool to obtain an algorithm for computation of, say, 
S-matrix elements.} The baggage we carry from Lorentz invariant,
perturbatively renormalizable, quantum field theory --- like the concepts of 
quanta, vacua, in-out states, Smatrix.... etc. --- is probably to be abandoned.  

Features (e) to (g) make  the situation worse. The fact that all matter gravitates [see (e)], once again stresses the need to abandon
description based on free field theory to handle virtual excitations with arbitrarily high energies.  An excitation with energy $E$ will probe length scales
of the order of $(1/E)$ and when $E\to E_P$, the  nonlinearity due to self gravity cannot be ignored. The same conclusion is applicable even to vacuum fluctuations of any field, including gravity. If we attempt to treat the ground state
of the gravitational field as the flat spacetime, we must conclude that the
spacetime structure at $L \la L_P$ will be dominated by quantum fluctuations of gravity and the smooth macroscopic spacetime can only emerge when the 
fluctuations are averaged over larger length scales.   

The difficulties mentioned above should caution one against approaching
the problem of quantum gravity as one of mathematics requiring a better
technical apparatus. There is 
very strong indication that the basic language of field theory
is inadequate to grapple with the complications introduced by gravity.
Perturbative language  which --- at best --- gives an algorithm to calculate
S-matarix elements, is not  going to be of much use in understanding the quantum structure of gravitational field. Most of the interesting questions --- possibly {\it all} the interesting questions --- in quantum gravity are non perturbative in character; whether a theory is perturbatively renormalizable or not
is totally irrelevent in this context. Conventional quantum 
field theory works  best when 
a static causal structure, global Lorentz frame, asymptotic in-out states,
bounded Hamiltonians and the language of vacuum state, particle excitations
etc., are supplied.  The gravitational field removes all these features, strongly hinting that we may be working
with an inadequate language. 
The gradual paradigm shift in the particle physics community from perturbative finiteness of supergravity (in early 80's) to  non perturbative description of superstrings (in late 90's)  represents
a grudging acceptance of the lessons from gravity. 
The history of these failures indicates that we have not been ruthless enough in attacking the problem.

\section{Quantum gravity from pure thought?}

Given the above results, is it possible to describe
the key features which must be present in any future, successful, theory of 
quantum gravity? I believe this can be done to certain extent
thereby providing some useful pointers. 

The fact that there will exist violent spacetime fluctuations at small scales
suggests that the macroscopic, continuum, description of spacetime can only be 
approximate and valid when quantum fluctuations are averaged over large
scales. The description of continuum spacetime in terms of, classical, Einstein's
equation is similar to the description of a solid by elastic constants or the description of a gaseous system by an equation of state. While the knowledge
of microscopic quantum theory of atoms and molecules will allow us, in principle, to construct the description in terms of elastic constants, the reverse 
process is unlikely to be unique. What one could hope is to take clues
from well designed thought experiments, thereby identifying some key 
generic features of the microscopic theory.

One might assume that the  microscopic description is in terms of certain [as yet unknown] variables $q_i$ and that the  conventional spacetime metric 
is obtained from these variables in some suitable limit. Such a process will necessarily involve coarse-graining over a class of microscopic descriptors of geometry.
I will now outline an argument which 
 suggests that there are {\it infinite} number of microscopic descriptors which are ``integrated out'' in proceeding from the fundamental description to spacetime description, \cite{tp98i}.  The argument proceeds in three steps: (1) Among all systems dominated by gravity, the universe
possess a very peculiar feature. If the conventional cosmological models are 
reasonable, then it follows that {\it our universe proceeded from quantum mechanical behaviour to classical behaviour in the course of dynamical evolution
defined by some intrinsic time variable}. It can be shown that a system with
bounded Hamiltonian can never make such a transition if classicality
is defined in terms of behaviour of a suitable Wigner function. It follows that the quantum cosmological description  of our universe, as a Hamiltonian system, should contain atleast one unbounded degree of freedom. It can also be shown that the unbounded mode --- which, in the case of FRW universe, corresponds to the expansion factor --- will go classical first, as is experienced in the evolution of the universe. (2) Let us next address the task of obtaining an unbounded Hamiltonian for an effective theory when the original theory contained a larger set of dynamical variables. It can again be shown that, if one starts with a bounded Hamiltonian for a system with finite  number of quantum fields and integrate out a   subset of them,
 the resulting Hamiltonian for the low energy theory cannot be unbounded. 
(3) Assuming that the original theory is describable in terms of a bounded Hamiltonian for some suitable variables, it follows that an infinite number of fields have to be involved
in its description and an infinite subset of them have to be integrated out
in order to give the standard low energy gravity. This feature is indeed present in one form or the other in the descriptions of quantum gravity based on
strings \cite{polchi96} or Ashtekar variables \cite{AA}.  My argument  suggests that this is indeed inevitable. 

If the description in terms of continuum spacetime is like theory of elasticity,
and we do not know the fundamental descriptors of spacetime, is there any way
of bridging the gap between the two? It turns out that this is  possible
by using the properties of macroscopic spacetime near the trapped surfaces.
I have given detailed arguments elsewhere  \cite{tp98ii} to show that the event horizon of a Schwarzschild blackhole acts as a magnifying glass, allowing us to probe
Planck scale physics. Consider, for example, a physical system described by a low energy Hamiltonian, $H_{\rm low}$. By constructing a blackhole made from the system with this Hamiltonian and requiring that the blackhole should have a density of states that is immune
to the details of the matter of which it is made, one can show
that the Hamiltonian, $H_{\rm true}$  describing the interactions of the system at transplanckian energies must be related to $H_{\rm low}$ by 
$ H_{\rm true}^2 = \alpha E_P^2 \ln [1 + (H^2_{\rm low} / \alpha E^2_p)]$
where $\alpha$ is a numerical factor. Of course, the description at transplanckian energies cannot be in terms of the 
original variables in the rigorous theory. The above formula should be interpreted as giving the mapping between an effective field theory (described by $H_{\rm true}$) and a conventional low energy theory (described by $H_{\rm low}$) such that the blackhole entropy will be reproduced correctly. 

In fact, one can do better and construct a whole class of effective field theories \cite{tp98iii} such that the one-particle excitations of these theories possess the same
density of states as a Schwarzschild blackhole. All such effective field theories are non local in character and possess a universal two-point function
at small scales. The nonlocality appears as a smearing of the fields over regions of the order of Planck length thereby confirming ones intuition 
about microscopic structures, trapped surfaces and blackhole entropy. 

If the physical description above Planck energy (or equivalently below Planck length) changes drastically,
 how can one modify the low energy description such that  the singularities in spacetimes and the perturbative divergences in quantum field theory are removed? This question cannot be answered rigorously without knowing the microscopic structure of spacetime. However two broad class of theories can be distinguished in terms of a general criterion. In the first class of theories, the low energy ($E\ll E_P)$ and high energy $(E\gg E_P)$ behaviour
are not related by any manner and the high energy sector of the theory {\it does}
affect the low energy behaviour significantly. If nature is built along these
lines, then we cannot predict much without knowing the full theory. On the
other hand, one can think of another class of theories in which the high energy
and low energy descriptions are related in a specified manner and are not completely
independent. The simplest form of such a relation will be a `duality' in
which the behaviour at a scale $E$ is related to a behaviour at scale $(E_P^2/E)$, or --- equivalently --- the behaviour at length scales $l$ and
$(L_P^2/l)$ are related. Implementing this duality in the path integral representation for a propogator, say, leads to a remarkable result \cite {tp97}: The effect of this duality is the same as assuming that the spacetime
possesses a `zero-point-length' and replacing the flat spacetime interval $(x-y)^2$ by $(x-y)^2+L_P^2$. I suspect that the converse is also true: if
the structure of the theory is such that planck length acts as a minimal
length to the spacetime, then the theory will possess a duality between length scales $l$ and
$(L_P^2/l)$. String theories do show related --- though not the same ---
features. If nature is built along these lines, then transplanckian physics
is dual to the low energy theory and must possess a description in terms of
some effective field theory. 

The key conclusions which emerge from all these are the following:
(i) It is unlikely that one will make genuine progress, unless the language
of quantum field theory is expanded to be capable of handling the
features listed in section 3. The question to understand is {\it not} why
gravity is difficult to quantise but {\it why the perturbative approach was
so unreasonably successful in dealing with other interactions ?}. This must be
because the conventional QFT is a wrong way of looking at physics though
it accidentally incorporated several features of the right [as yet unknown] approach ---
as was in the case of, say, old quantum theory. Rethinking about QED in a possibly
new language might offer hints on how to proceed further. (ii) Given the 
unlikely event of experimental confirmation of quantum gravity, it is necessary
to attempt a top-down approach [ classical gravity $\to$ effective field
theory $\to$ microscopic spacetime descriptors ] using, say, well-defined thought experiments.
In this regards, spacetimes with trapped surfaces will be valuable. (iii) It is
also important to worry, at a conceptual level, the effect of transplancian
physics on low energies. If this effect is not to be unreasonably strong, thereby
killing predictability, it is necessay that the low energy theory is
protected by some kind of
duality mapping relating transplancian energies to low energies. A progam for
quantum gravity, along these lines, holds promise.

I thank Apoorva Patel for several illuminating discussions.

\end{document}